\documentclass[10pt,letterpaper]{scrartcl}
\usepackage[utf8]{inputenc}
\usepackage[english]{babel} 
\usepackage{siunitx}
\usepackage[paper=letterpaper,left=25mm,right=25mm,top=25mm,bottom=30mm]{geometry}
\usepackage[round]{natbib}
\usepackage[colorinlistoftodos]{todonotes}
\usepackage{dblfloatfix}
\usepackage{float}
\usepackage[breaklinks, colorlinks=true, citecolor=black, urlcolor=black, linkcolor=black]{hyperref}

\usepackage{authblk}
\usepackage{multicol}
\usepackage{graphicx}

\usepackage{amsmath}
\usepackage[format=plain,labelfont=bf,up,textfont=it,up,small]{caption}
\usepackage{sidecap}

\usepackage[bottom, hang]{footmisc}
\usepackage{scrpage2}
\definecolor{boxcolor}{HTML}{e3e9f2}
\definecolor{boxcolor2}{HTML}{f2f2f2}
\usepackage{nameref}

\title{\vspace{-1.1cm}\LARGE Combining crowd-sourcing and deep learning to explore the meso-scale organization of shallow convection}

\author{\large Stephan Rasp\thanks{Technical University of Munich, Germany. Corresponding author: stephan.rasp@tum.de}\  \, \, Hauke Schulz\thanks{Max Planck Institute for Meteorology, Hamburg, Germany}\, \, \, Sandrine Bony\thanks{Sorbonne Universit\'e, LMD/IPSL, CNRS, Paris, France}\, \, \, Bjorn Stevens\textsuperscript{$\dagger$}}
\date{\vspace{-8ex}}

\begin{document}

\maketitle

\begin{abstract}
Humans excel at detecting interesting patterns in images, for example those taken from satellites. This kind of anecdotal evidence can lead to the discovery of new phenomena. However, it is often difficult to gather enough data of subjective features for significant analysis. This paper presents an example of how two tools that have recently become accessible to a wide range of researchers, crowd-sourcing and deep learning, can be combined to explore satellite imagery at scale. In particular, the focus is on the organization of shallow cumulus convection in the trade wind regions. Shallow clouds play a large role in the Earth's radiation balance yet are poorly represented in climate models. For this project four subjective patterns of organization were defined: Sugar, Flower, Fish and Gravel. On cloud labeling days at two institutes, 67 scientists screened 10,000 satellite images on a crowd-sourcing platform and classified almost 50,000 mesoscale cloud clusters. This dataset is then used as a training dataset for deep learning algorithms that make it possible to automate the pattern detection and create global climatologies of the four patterns. Analysis of the geographical distribution and large-scale environmental conditions indicates that the four patterns have some overlap with established modes of organization, such as open and closed cellular convection, but also differ in important ways. The results and dataset from this project suggests promising research questions. Further, this study illustrates that crowd-sourcing and deep learning complement each other well for the exploration of image datasets.
\end{abstract}

\begin{figure*}[ht!]
\noindent\colorbox{boxcolor}{%
	\parbox{\linewidth}{%
\centering \large \textsf{
    Crowd-sourcing and deep learning are combined to explore the meso-scale organization of shallow clouds in the subtropics.}
    }%
}
\end{figure*}\vspace{-2ex}

\begin{multicols}{2}

\subsection*{Introduction}

A quick glance at an image, be it taken from a satellite or produced from model output, is often sufficient for a scientist to identify features of interest. Similarly arranged features across many images form the basis for identifying patterns. This human ability to identify patterns holds true also in situations where the features, let alone the patterns they build, are difficult to describe objectively---a situation which frustrates the development of explicit and objective methods of pattern identification. In these situations, machine learning techniques, particularly deep learning (see Sidebar 1), have demonstrated their ability to mimic the human capacity for identifying patterns, also from satellite cloud imagery \citep[e.g.,][]{Wood2006}.  However, the application and assessment of such techniques is often limited by the tedious task of obtaining sufficient training data, so much so that (in cloud studies at least) these approaches have not been widely adopted, let alone assessed.

Recently, \cite{Stevens2019a} described a collective cloud classification activity by a team of 13 scientists supported by the International Space Science Institute (ISSI).  This ISSI team aimed to identify mesoscale cloud patterns in visible satellite imagery taken over a trade-wind region east of Barbados. Organization, or clustering, of clouds has been shown to have important implications for climate in the case of deep convection \citep{Tobin2012}, which raises the question to what extent this is the case in shallow clouds. The ISSI-team's hand-labeling effort resulted in around 900 subjectively classified images. An initial application of machine learning to these images (by the first author) proved promising but also highlighted the need for more training data in order to obtain robust and interpretable results.

Based on these first insights the authors organized a crowd-sourced project (see Sidebar 2) that would allow us to collect a substantially larger set of labeled images. This activity was designed to provide a better foundation for the application of machine learning to the classification of patterns of shallow clouds, as well as to explore methodological questions raised when attempting to marry crowd-sourcing with machine learning to address problems in climate and atmospheric science. Specifically we sought to answer four questions:
\begin{description}
    \item[Q1] How should a community-driven labeling exercise be set up to ensure a) a good user experience for participants and b) the usefulness of the gathered data for subsequent analysis?
    \item[Q2] Can a diverse set of scientist identify the subjective modes of cloud organization established by the ISSI team with satisfactory agreement to warrant further scientific analysis?
    \item[Q3] Can a deep learning algorithm learn to classify images as well as trained scientists?
    \item[Q4] To the extent that a machine can be trained to classify large numbers of images, what can be learned from applying this algorithm to global data?
\end{description}

In this paper, we present our findings. They suggest that, for suitable problems, the combination of crowd-sourcing and deep learning allow scientists to analyze data on a scale beyond what would be possible with traditional methods. Though our main findings will be of particular interest to researchers interested in the mesoscale organization of shallow clouds, the methods used to obtain them may be of more general interest, and are presented with this in mind. 

We begin by describing how the cloud patterns (or classes) we sought to classify were defined, followed by a summary of the crowd-sourcing project. Then the results from the human data is presented before we explain how deep learning is used to extend the analysis. Finally, we summarize our findings as pertains to the above stated research questions, from which inferences of potential relevance to future studies are drawn.

\begin{figure*}[ht!]
	\noindent
	\includegraphics[width=1\linewidth]{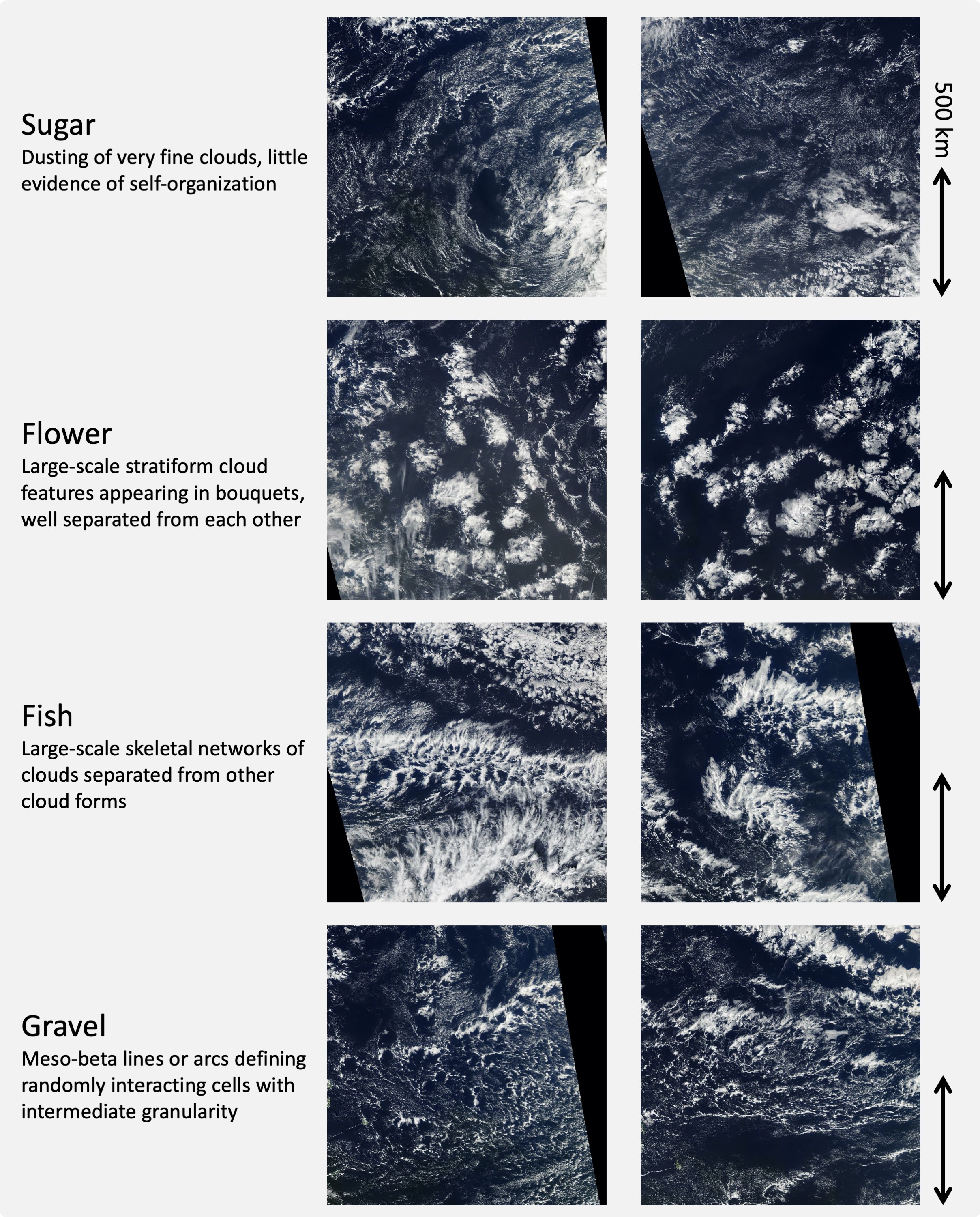}
	\caption{Canonical examples of the four cloud organization patterns as selected by the ISSI team.}
	\label{fig:1}
\end{figure*}

\begin{figure*}[hb!]
\centering
\fbox{%
	\parbox{0.9\linewidth}{
	\large \centering{\textbf{Box 1: Deep learning for vision tasks in the geosciences}}\\
	\small \justify Deep learning describes a branch of artificial intelligence based on multi-layered artificial neural networks \citep{Nielsen2015}. In recent years, this data-driven approach has revolutionized the field of computer vision which up to 2012 was to a large extent based on hard-coded feature engineering \citep{LeCun2015}. More specifically, the success of deep learning in vision tasks is based on convolutional neural networks which exploit the translational invariance of natural images (i.e. a dog is a dog whether it is in the top right or bottom left of the image) to greatly reduce the number of unknown parameters to be fitted. Deep neural networks also have many potential applications in the Earth sciences, particularly where already existing deep learning techniques can be transferred to geoscientific problems \citep{Reichstein2019}. A perfect example of this is the detection of features in images, the topic of this study. 
	One obstacle is that deep learning requires a large number, typically several thousands, of hand-labeled training samples. For Earth science problems, these are usually not available. For this reason, previous studies that used deep neural networks to detect atmospheric features relied on training data created by traditional, rule-based algorithms \citep{Racah2016, Liu2016, Hong2017a, Kurth2018, Mudigonda2017}. A notable exception is the aforementioned study by \cite{Wood2006}.  They hand-labeled 1000 images of shallow clouds and used a neural network to classify them into four cloud types, making it a predecessor to our study. 
	}%
}
\end{figure*}

\section*{Sugar, Flowers, Fish and Gravel}

Mesoscale patterning of shallow cumulus is a common feature in satellite imagery. However, organization on these scales is largely ignored in modeling studies of clouds and climate. This applies to process studies with large-eddy simulations \cite[e.g.,][]{RieckEtAl2012,Bretherton2015b} as well as general circulation models, be it in traditional or super-parameterizations \citep{Arakawa1974,ParishaniEtAl2018}.

The prevalence of mesoscale patterning in satellite cloud imagery led the ISSI team \citep{Stevens2019a} to identify four cloud patterns that frequent the lower trades of the North Atlantic.  They named these patterns Sugar, Flower, Fish and Gravel (Fig.~\ref{fig:1}). The choice of new and evocative names was motivated by the judgement that the patterns were different than those that have been previously described, for instance in studies of stratocumulus or cold-air outbreaks.  Support for this judgement is provided by an application of the neural network from \cite{Wood2006} and \cite{Muhlbauer2014}, which was trained to distinguish between ``No Mesoscale Cellular Convection (MCC)'', ``Closed MCC'', ``Open MCC'' and ``Cellular, but disorganized''.  When applied to the scenes classified by the ISSI team the algorithm mostly resulted in the ``disorganized'' classification (Personal communication with I. L. McCoy). Despite the lack of a simple link between the patterns classified by the ISSI team and patterns previously described in the literature, below we point out previously identified patterns that may be related to the four patterns used here.

``Sugar'' describes wide-spread areas of very fine cumulus clouds. Overall these fields are not very reflective, do not have large pockets of cloud-free regions and, ideally, exhibit little evidence of meso-scale organization. Often, though, they are embedded within the larger-scale flow which gives them some structure. In strong flow, Sugar can form thin "veins", or feathers, which have previously described as dendritic clouds \citep{Nicholls2007}.

``Flowers'' are areas with isotropic cloud structures, each ranging from 50 to 200 km in diameter, with similarly wide cloud-free regions in-between. This pattern overlaps to some degree with canonical closed-cell MCC. Flowers, however, are often less densely packed than typical closed cells, which only have narrow cloud-free regions at the edges, and they are identified well outside of regions where stratocumulus are found \citep{Norris1998}. One hypothesis is that they are successors of more closely packed closed-cell MCC which are in the process of breaking up.

``Fish'' are elongated, skeletal structures that sometimes span up to 1,000 km, mostly longitudinally. As noted by \cite{Stevens2019a}, these features appear similar to what \cite{GarayEtAl2004} called actinoform clouds.  They presented examples of these particularly well structured cloud forms taken from all ocean basins, near but typically downwind of regions where stratocumulus maximize. To the extent Fish are variants of the actinoform clouds found by \citeauthor{GarayEtAl2004}, they may be more common than previously thought.

Finally, ``Gravel'' describes fields of granular features marked by arcs or rings. The typical scale of these arcs is around 20\,km. We suspect that these patterns are driven by cold pools caused by raining cumulus clouds \citep{Rauber2007}. In this regard, Gravel is fundamentally different from open-cell MCC, which has larger cells that are driven by overturning circulations in the boundary layer. However, the line between these two mechanisms can blur at times. 

It is also interesting to compare our subjectively chosen labels to those of \cite{Denby2020} who used an unsupervised learning algorithm to automatically detect different types of cloud organization (their Fig. 2). Some of their patterns bear resemblance to our classes, e.g. "Sugar" seems to most closely correspond to their patterns A and B, "Gravel" to G and H. However, their automatically detected classes appear less striking to the human eye.

\section*{Crowd-sourced labels}

To obtain a large pool of labeled images from the community, an accessible user interface is needed. Zooniverse\footnote{\url{https://www.zooniverse.org/projects/raspstephan/sugar-flower-fish-or-gravel}} is an open web platform that enables researchers to organize and present research questions in ways that enable contributions from the broader public (see also Sidebar 2). Its flexibility in serving and presenting images, choosing between different labeling tasks and its ability to monitor and organize the information associated with the labeling activities made Zooniverse very well suited for our task.

\begin{figure*}[ht!]
	\noindent
	\includegraphics[width=1\linewidth]{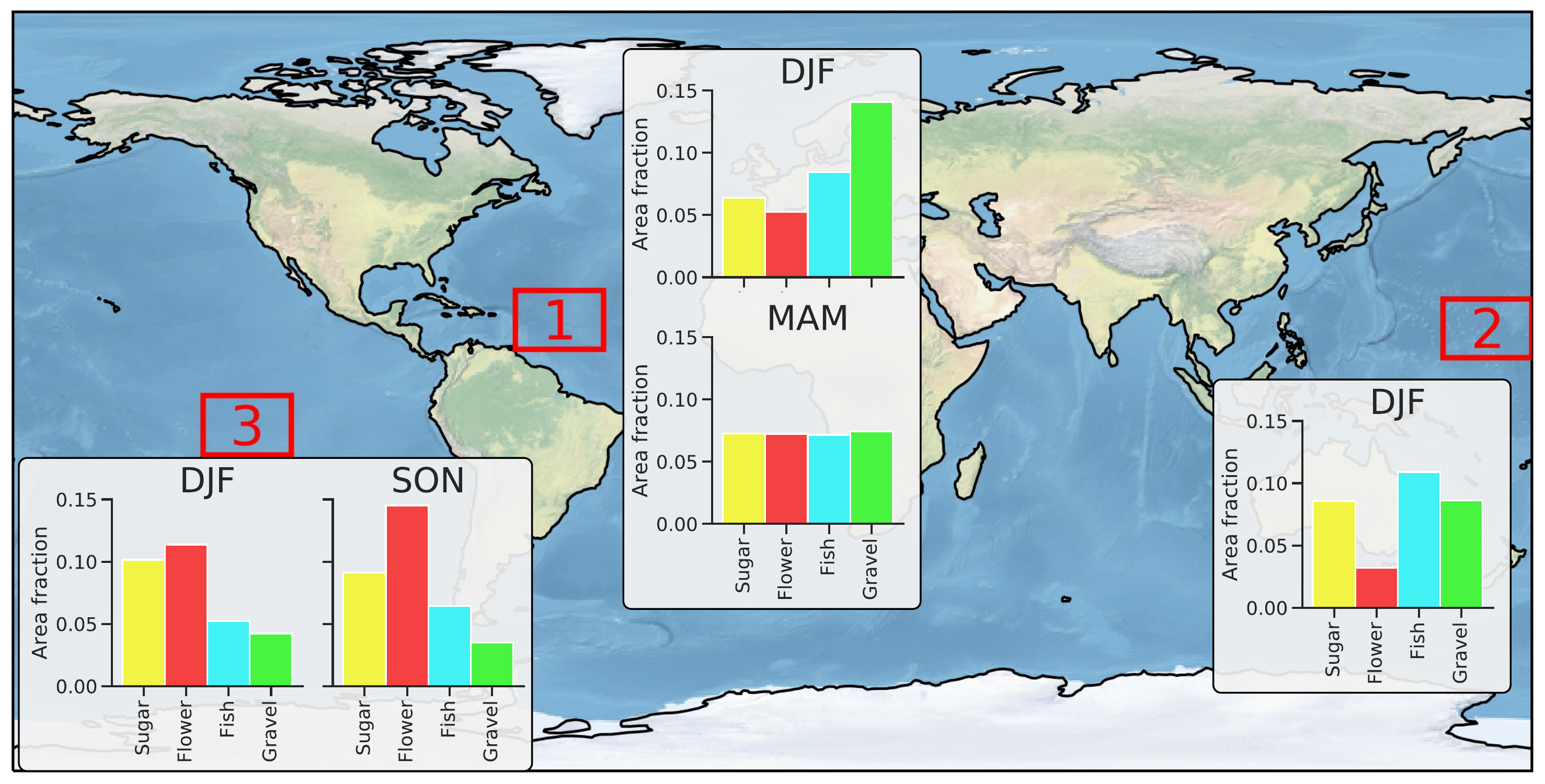}
	\caption{World map showing the three regions selected for the Zooniverse project. Bar charts are showing which fraction of the image area was classified into one of the four regions by the human labelers. Note that the areas do not add up to one. The remaining fraction was not classified.}
	\label{fig:2}
\end{figure*}

For our project we downloaded roughly \num{10000} $\ang{14}\times\ang{21}$ (lat-lon) Terra and Aqua MODIS visible images from NASA Worldview. To select the regions and seasons, we started with the Boreal winter (DJF) east of Barbados as a reference. Barbados is home to the Barbados Cloud Observatory \citep{Stevens2016}. The clouds in its vicinity were not only the focus of the ISSI-teams study, but have more generally come to serve as a laboratory for studies of shallow clouds and climate \citep{Stevens2016,MedeirosNuijens2016,Stevens2019a,Bony2017}.  To obtain more images and sample a greater diversity of clouds we subsequently added images from two further regions in the Pacific, which were chosen based on their climatological similarity to the original study region upwind of Barbados (Fig.~\ref{fig:2}; see Supplement for details). Images were downloaded for an eleven year period from 2007 to 2017.  

\begin{figure*}[hb!]
\centering
\fbox{%
	\parbox{0.9\linewidth}{
	\large \centering{\textbf{Box 2: Crowd-sourcing}}\\
	\small \justify Crowd-sourcing describes projects where a task is collaboratively solved by a group of people. This can be a small research group or a large group of internet users. One of the first examples of crowd-sourcing in the natural sciences is Galaxy Zoo\footnote{\url{https://www.zooniverse.org/projects/zookeeper/galaxy-zoo}}, a project that has citizen scientists classify different galaxy types and has produced 60 peer-reviewed publications so far. An early meteorological example focused on estimating hurricane intensity \citep{Hennon2015}. Recent climate projects on the crowd-sourcing platform Zooniverse (\url{https://www.zooniverse.org/projects/edh/weather-rescue}, \url{https://www.zooniverse.org/projects/drewdeepsouth/southern-weather-discovery}) asked volunteers to transcribe old, hand-written weather records. Thanks to the collaboration of many individuals such projects produce a wealth of data that would be unattainable for a single scientist. Note that for this paper we understand the term crowd-sourcing to indicate active labor by the participants rather than providing data through personal sensors or cameras. For a broader review of citizen science and crowd-sourcing studies in the geosciences, see \cite{Zheng2018}.  
	}%
}
\end{figure*}

\begin{figure*}[ht!]
	\noindent
	\includegraphics[width=1\linewidth]{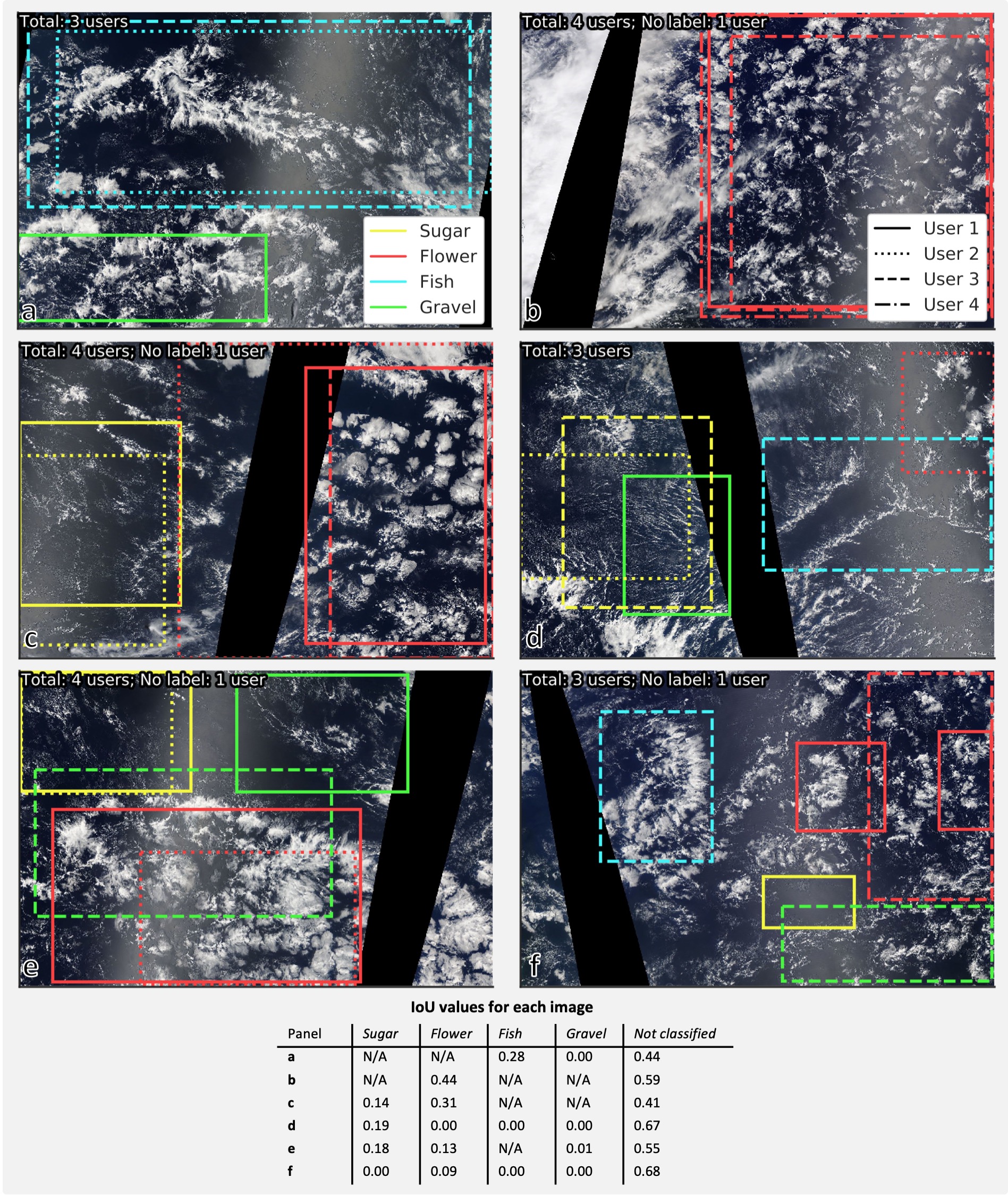}
	\caption{Six example images showing annotations drawn by human labelers. Different line styles correspond to different users. In addition the IoU values for each image and class are shown in the table.}
	\label{fig:3}
\end{figure*}

\cite{Stevens2019a} speculated that their protocol of assigning a single label to the entire $\ang{10}\times\ang{20}$ image resulted in considerable ambiguity and disagreement between labelers. In an attempt to minimize this issue we presented participants with slightly larger images and experimented with ways to allow the labeling of multiple, and possibly overlapping sub-regions.  This was accomplished by allowing users to draw rectangles around regions where they judged one of the four cloud patterns to dominate (see Fig.~\ref{fig:3} for examples). Participants had the possibility to draw any number of boxes, including none, with the caveat that the box would cover at least 10\% of the image.  We arrived at this setup after experimenting with other options, such as labeling subsections of a predefined grid, or allowing users to label regions that they defined using polygons with an arbitrary number of sides.  We opted for the rectangles to increase labeling speed and improve the user experience.  Our thinking was that it would be better to have less accurate but more plentiful data, and that given the vague boundaries of the cloud structures, it was anyway doubtful that a more accurate labeling tool would add much information.  As we will show later, this thinking paid off for the machine learning models we trained.

The Zooniverse interface was further configured to serve participants an image randomly drawn from our library of \num{10000} images. After being classified by four different users, images were retired, i.e. removed from the image library. In addition, no user was shown the same image twice.  
With the interface in place, cloud classification days were set up at the Max Planck Institute for Meteorology in Hamburg, Germany on Nov 2nd and at the Laboratoire de M\'et\'eorologie Dynamique in Paris, France on Nov 29th, 2018. After a brief instruction at the start of the day and a warm-up on the training dataset, 67 participants, most of them researchers, from the two institutes, labeled images for an entire day.  The activity yielded roughly \num{30000} classified images, i.e., each image was classified about three times on average.  Because an image could have sub-regions with different classifications, the number of labels was somewhat larger with \num{49000}. On average, participants needed around \SI{30}{s} to classify one image, amounting to approximately \SI{250}{\hour} of concentrated human labor. There was however considerable differences among users, as the interquartile range in classification times ranged from \SI{20}{s} to \SI{38}{s}. Overall, the four patterns occupied similarly large areas but notable differences occurred depending on the geographic region and season (Fig.~\ref{fig:2}).

\section*{Inferences from human labels}

Given the subjective nature of labels assigned by visual inspection, our first research question was to what extent the human labelers agreed with each other. In the initial classification exercise of 900 images reported in \cite{Stevens2019} a majority of scientists agreed on one pattern in \SI{37}{\percent} of the cases, significantly more than random. In this project, in addition to choosing the category of the clouds, participants also had to choose the location. To explore the agreement we started by looking at many examples, six of which are reproduced in Fig.~\ref{fig:3}. Many more can be found at \cite{human}. The most notable conclusions from this visual inspection are that users agreed to a high degree on features that closely resemble the canonical examples of the four classes but also that there was a lot of disagreement otherwise. Take Fig.~\ref{fig:3}a where two out of three participant agreed on the presence of Fish in the top half of the image; or Fig.~\ref{fig:3}b where three out of four participants recognized a region of Flowers. On the other hand Fig.~\ref{fig:3}d shows an example of an image with plenty of ambiguity. Also note that users applied different methodologies when labelling, some labeling a single large region, other many small regions. Overall, we came to the conclusions that, while certainly noisy, clear examples of what was defined as Sugar, Flower, Fish and Gravel could be robustly detected.

\begin{figure*}[ht!]
	\noindent
	\includegraphics[width=1\linewidth]{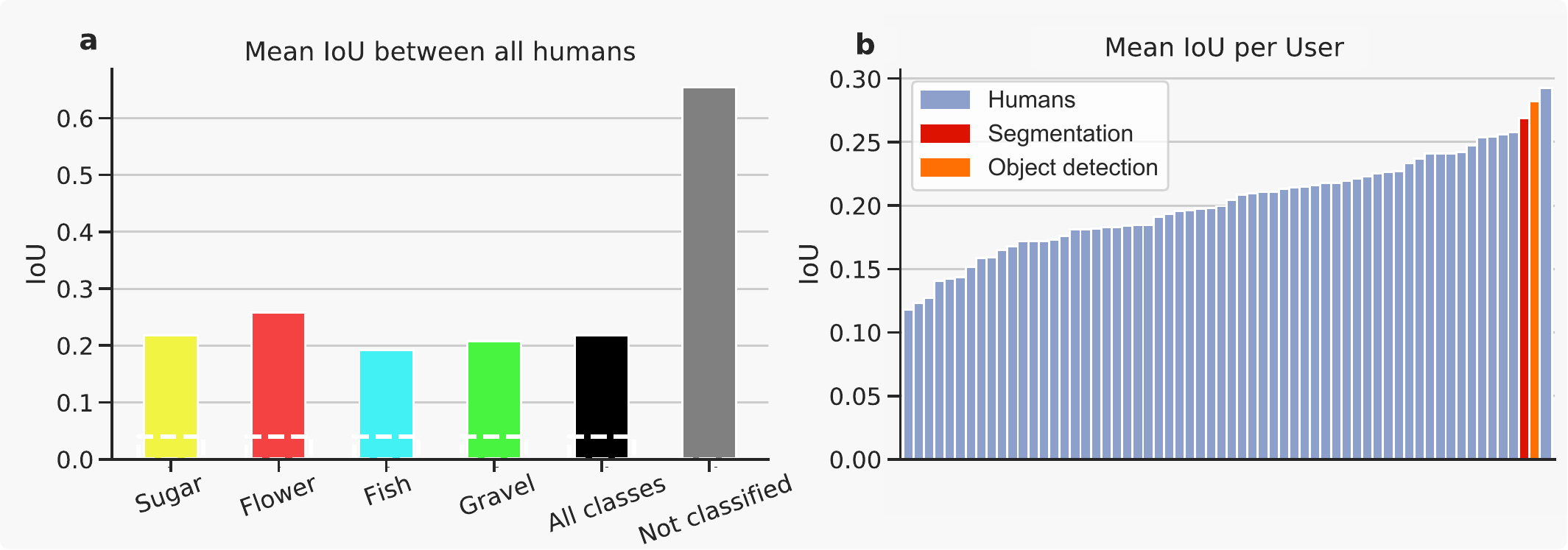}
	\caption{(a) Mean IoU between humans. Dashed line represents random IoU; see text for details. (b) Mean IoU for each human participant and the two deep learning algorithms for a validation dataset.}
	\label{fig:4}
\end{figure*}

Next, we aimed to quantify the agreement. To our knowledge there is no standard way of evaluating subjective labels from multiple users. The most commonly used metric for comparing a label prediction with a ground truth is the Intersect over Union (IoU) score, also called the Jaccard index. Given two sets, $A$ and $B,$ it is defined as the ratio of their intersection to their union, i.e., $I = A \cap B$ divided by $U = A \cup B$. An IoU score of one indicates perfect overlap, while zero indicates no overlap. We adapted the IoU score to this task by first iterating over every image and then computing the intersect $I_\text{image}$ and union $U_\text{image}$ for every user-user combination for this image. To compute the final "Mean IoU between humans" we computed the sum of the intersect and union over all images: $I = \sum_\text{image} I_\text{image}$ and $U = \sum_\text{image} U_\text{image}$. This was done for each cloud class separately. We also computed an IoU score for the "Not classified" area. Finally, the "All classes" IoU was computed by additionally taking the sum of $I$ and $U$ over all classes. The results for the inter-human mean IoU are shown in Fig.~\ref{fig:4}a.

At first glance, IoU values of around 0.2 seem low. In classical computer vision tasks such values would certainly indicate low agreement. However, as mentioned above, this dataset is different from classical object detection tasks in that there are more than two labelers for most images and there are many cases in which one or more participants did not label an image. In fact, the primary reason for the low mean IoU score are zero values, which arise from some users detecting a feature while others did not. Take Fig.~\ref{fig:3}b as an example. Here, three of four users agreed to a high degree of accuracy on the location of Flowers but the last user did not submit a label. This results in three "no label"-"label" comparisons and three "label"-"label" comparisons. Even with perfect agreement between the three Flower labelers, the mean IoU would only be 0.5. In reality it is 0.44 for this example. These "no label"-"label" pairs with IoU = 0 make up 63\% of all user-user comparisons (see Supplement Fig.~S2). Omitting these gives a mean IoU of 0.43. To get a feel for what this value means consider the two Sugar rectangles in Fig~\ref{fig:3}d, which have an IoU of 0.46. The table at the bottom of Fig.~\ref{fig:3} shows the mean IoU values for each of the example images. These numbers suggest that even for images where one would visually detect a high degree of agreement between the users, the IoU numbers are quite low. For this reason, the actual values should not be compared to other tasks where the IoU is used. Rather, for this paper they simply serve the purpose of comparing different classes and methods. To further illustrate this point we computed the IoU score for many randomly drawn labels from the same number and size distribution as the human labels, which gives an IoU of only 0.04. What the numbers do show is that there are noticeable differences between the four patters. People agreed most on Flowers while Fish proved more controversial. With regards to Q2, we came to the conclusion that, despite the noise in the labels, there was sufficient consensus between the participants on clear features to warrant further analysis, especially since as we will see the noise will largely disappear in the statistical average.

Another question that the new methodology of labeling allows us to answer is whether or not the patterns tend to span larger or smaller areas. Based on the Zooniverse labels, Flower boxes tended to be largest, covering around \SI{25}{\%} (around 900,000\,km$^2$) of the image. Fish and Gravel were somewhat smaller with a box size of around \SI{20}{\%} (around 720,000\,km$^2$). Sugar spanned regions smaller yet with boxes only taking up \SI{15}{\%} (around 540,000\,km$^2$) of the image on average. Because the initial classification by \cite{Stevens2019a} required labelers to identify the entire scene, the relative infrequency with which they detected Sugar is likely due in part to the infrequency with which it covers large areas.

\begin{figure*}[ht!]
    \noindent
    \includegraphics[width=1\linewidth]{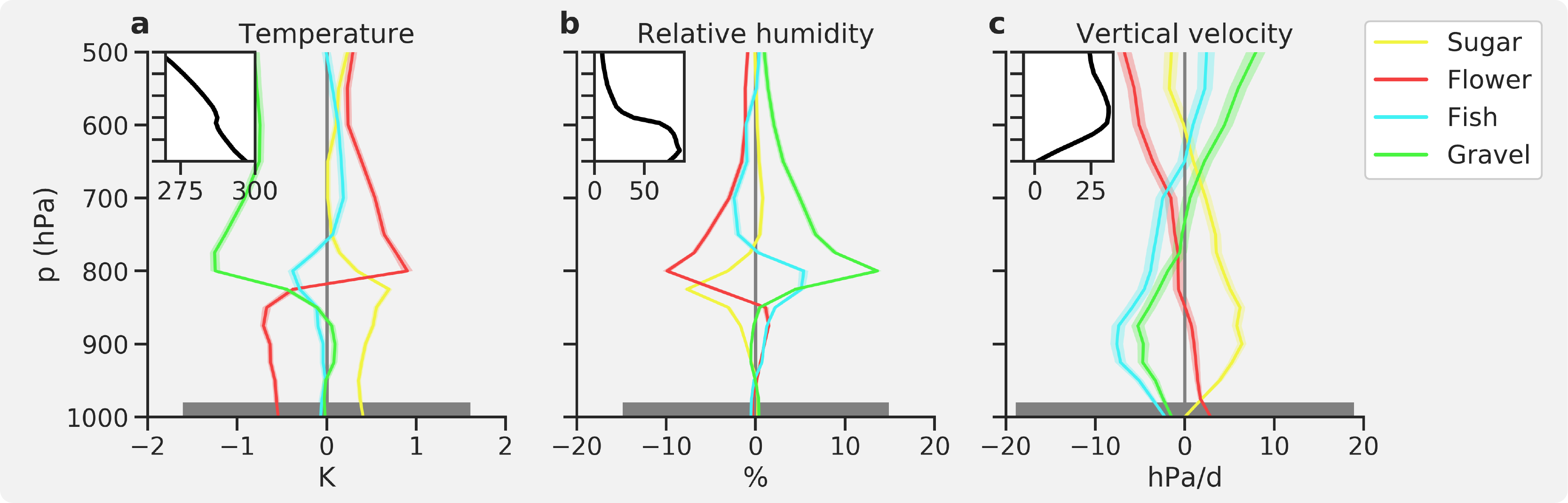}
    \caption{Median of large-scale environmental conditions corresponding to the four patterns as identified by the human labelers. Figures show deviations of (a) temperature, (b) specific humidity and (c) vertical velocity (shown in $\omega = \mathrm{d}p/\mathrm{d}t$) relative to the climatological mean which is shown in the figure insets. The shading about the lines shows the standard error, and hence the statistical difference between the mean conditions associated with any particular pattern.  The bar at the base of the figure shows the average inter-quartile spread (for the level where this spread maximizes, around \SI{800}{\hecto\pascal}) in the thermodynamic state associated with each pattern, indicating that the conditions associated with any given pattern can vary considerably.}
    \label{fig:5}
\end{figure*}

Further we can ask whether the four patterns, which were purely chosen based on their visual appearance on satellite imagery, actually correspond to physically meaningful cloud regimes. To investigate this, we created composites of the large-scale conditions from ERA-Interim reanalyses\footnote{\url{https://www.ecmwf.int/en/forecasts/datasets/reanalysis-datasets/era-interim}} corresponding to each pattern (Fig.~\ref{fig:5}). To the extent the ERA-Interim accurately represents the meteorological conditions in the region, the composites suggest that Sugar, Flower, Fish and Gravel appear in climatologically distinct environments. This is supported by the standard error being smaller than the difference between the patterns. The standard error is a measure for how well the mean conditions of a given pattern can be estimated and is defined as $\sigma / \sqrt{N}$, where $\sigma$ is the standard deviation and $N$ is the sample size. At the same time, there is variability between individual profiles within a composite, as shown by the inter-quartile range. Hence, while the compositing suggests that the occurrence of a particular pattern is associated with significant changes in the large-scale environmental conditions, this is clearly not the only factor at play, and things like airmass history are likely also important.

Flowers tend to be associated with a relatively dry and cold boundary layer with a very strong inversion (note that Fig.~\ref{fig:5} shows deviations from the climatological mean). Sugar on the other hand appears in warm and humid boundary layers with strong downward motion maximizing near the cloud base. For Fish and particularly Gravel, on the other hand, the inversion and downward motion is rather weak. The fact that Flowers and Gravel are essentially opposites in terms of their environmental profiles suggests that they are not simply manifestations of closed and open-cellular convection, which often transition smoothly into one another in similar large-scale environments \citep{Muhlbauer2014}.

\section*{Application of deep learning}

While the \num{10000} images labeled on Zooniverse already provide a useful dataset for further analysis, they only cover a small fraction of the globe for a small fraction of time. Only \SI{0.6}{\%} of the data available during the selected eleven year period were labeled. In this section, we explore whether deep learning (see Sidebar 1) can help to automate the detection of the four organization patterns and if so what can be learned from it.

\begin{figure*}[ht!]
	\noindent
	\includegraphics[width=1\linewidth]{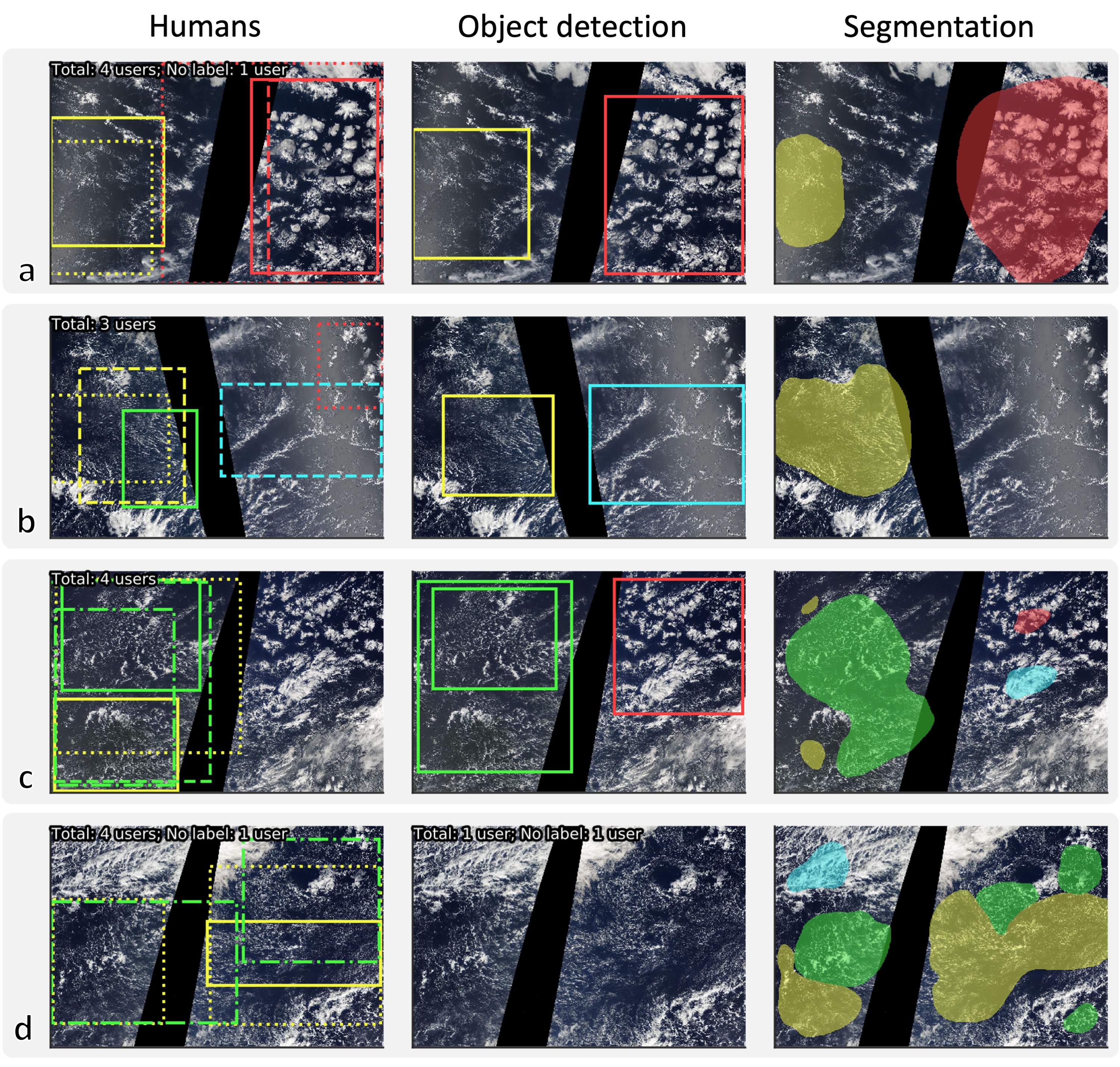}
	\caption{Human and machine learning predictions for four images from the validation set. Note that images a) and b) are also shown in Fig.~\ref{fig:3}.}
	\label{fig:6}
\end{figure*}

The pattern recognition task can be framed as one of two machine learning problems: object detection and semantic segmentation. Object detection algorithms draw boxes around features of interest, essentially mirroring what the human labelers were doing. In contrast, segmentation algorithms classify every pixel of the image. Fig.~\ref{fig:6} shows examples of these two approaches for images from a validation dataset that was not used during training (see \cite{ml} for more, randomly chosen examples). Details about the neural network architectures and preprocessing steps can be found in the Supplement. Both types of algorithm accurately detect the most obvious patterns in the image and agree well with human labels. Neither algorithm is perfect, however. The object detection algorithm sometimes misses features, as is visible in Fig.~\ref{fig:6}d. The segmentation algorithm, on the other hand tends to produce relatively small patches (Fig.~\ref{fig:6}c and d) because, other than humans and the object detection algorithm, in which the range of possible box sizes is an adjustable parameter, it has not been given instructions to only label larger patches. An interesting and advantageous feature of the segmentation algorithm is that, despite all training labels being rectangular, it appears to focus on the actual, underlying shape of the patterns, as visible by the rounded outlines of the predicted shapes. This suggests that despite the uncertainty in the human dataset, the deep learning algorithms are able to filter out a significant portion of this noise and manage to distill the underlying human consensus.

To quantitatively compare the deep learning algorithms against the human labelers, we compute the mean IoU for each human individually as well as for the two algorithms (Fig.~\ref{fig:4}b). Both algorithms show a large agreement with the human labels for a random validation dataset. The fact that the scores are higher than the mean inter-human IoU directly reflects the fact that the algorithms tend to produce less noisy predictions. Further analysis shows that the algorithms inherit some biases from the human training labels. The frequency and accuracy of the predicted labels is higher for patterns with a higher inter-human agreement, most notably flowers (Supplemental Fig.~3), which could slightly bias the deep learning predictions.

The main advantage of deep learning algorithms is that they are very fast at inference, one second per image compared to the 30 seconds a human needed on average, and they are more scalable.  This allows us to apply the algorithm to the entire globe (Fig.~\ref{fig:7}a; see Supplement for details). A healthy skepticism is warranted  when applying machine learning algorithms outside of their training regime \citep{Rasp2018c, Scher2019}. A visual inspection of the global maps (see \cite{global} for more examples), however, suggests that the algorithm's predictions are reasonable and physically interpretable as discussed below. Naturally, over land the predictions have to be assessed with greater care because no land was present in the training dataset. Nevertheless, Fig.~\ref{fig:7}a suggests that the algorithm even appears to correctly identify shallow cumuli over the tropical landmasses as sugar. 

To obtain global climatologies of Sugar, Flower, Fish and Gravel we ran the algorithm on daily global images for the entire year of 2017 (Fig.~\ref{fig:7}b--e). The resulting heatmaps reveal coherent hotspots for the four cloud patterns.  The spatial distribution of these hotspots helps answer some further questions raised by the ISSI team's study. For instance, the heat maps indicate that organization is most common over the ocean.  Only Sugar -- the one pattern characterized by its lack of mesoscale organization -- was identified over land (but keeping in mind the potential bias of the algorithm).  Our results also indicate that Sugar, followed by Flower, are the most common forms of organization globally. This indicates a bias arising from the ISSI team's focus on a single study region, as large areas of Sugar are relatively rare near Barbados.  A prevalence of Sugar in the trades adjacent to the deep tropics, and regions such as the Arabian sea, is consistent with its coincidence in association with strong low-level subsidence and a somewhat drier cloud layer (as seen by the large-scale composites, Fig.~\ref{fig:5}) indicating that it might be most favored in regions where convection is suppressed by strong subsidence from neighboring regions of active convection, or strong-land sea circulations.

Flowers prevail slightly downstream of the main stratocumulus regions.  Composites of the environmental conditions in which they form show them to be, on average, associated with large scale environmental conditions characterized by more pronounced lower tropospheric stability, and a somewhat drier free troposphere (Fig.~\ref{fig:5}).  This lends credence to the idea that they are manifestations of closed-cell MCCs.  Whereas the climatology of closed-cell MCC by \cite[][their Fig.~5]{Muhlbauer2014} shows similar hotspots in the subtropics, it also has strong maxima across the mid and high-latitude oceans. The absence of such hotspots in our classification of Flowers could indicate a bias of our algorithm towards the regions it was trained on. However, it could also suggest that ``Flowers'' differ from typical closed-cellular convection in their scale and spacing. 

\begin{figure*}[ht!]
    \noindent
    \includegraphics[width=1\linewidth]{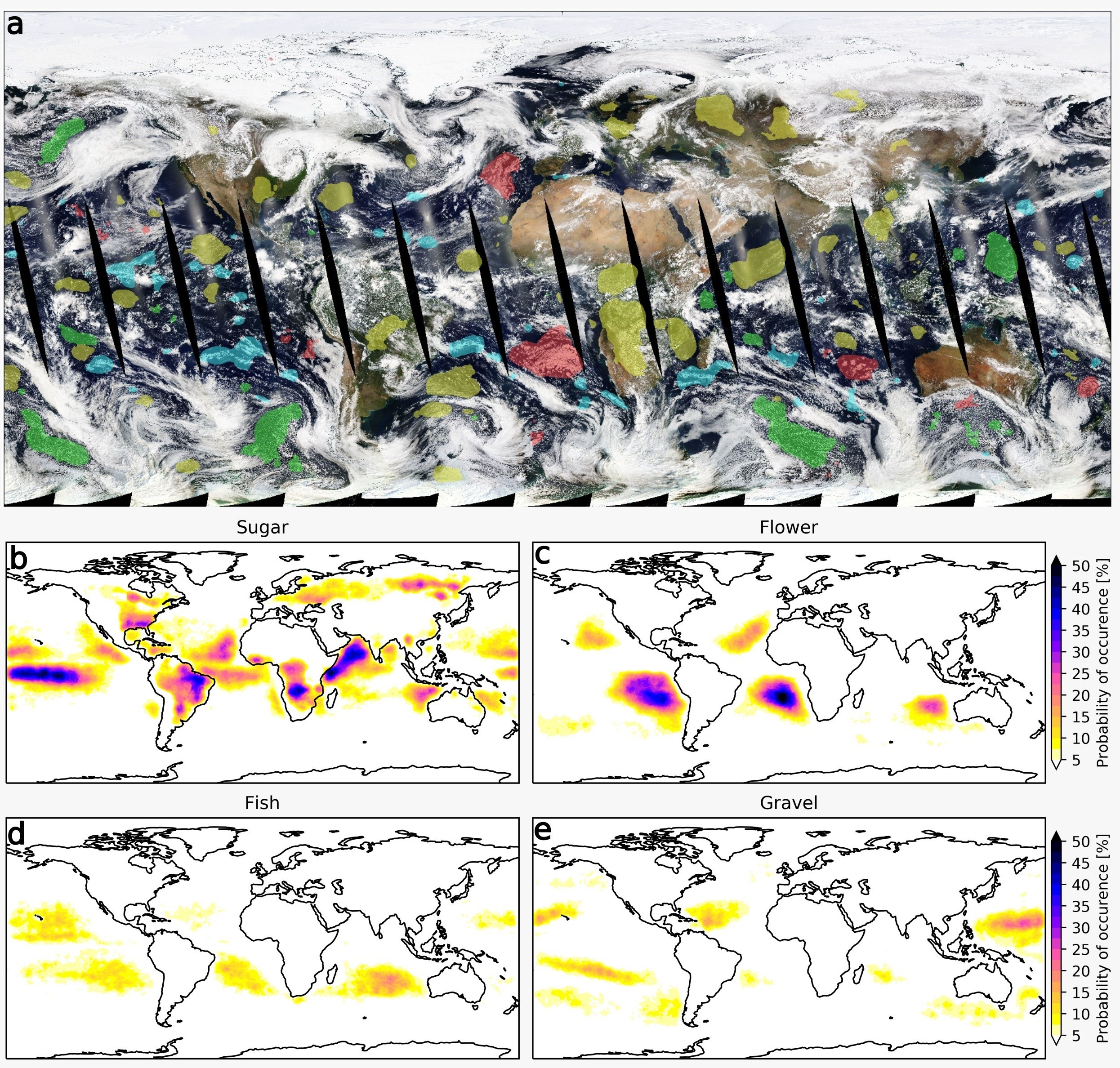}
\caption{(a) Global predictions of the image segmentation algorithm for May 1 2017. The colors are the same as in the previous figures. See \cite{global} for more examples. (b--e) Heatmaps of the four patterns for the year 2017.}
    \label{fig:7}
\end{figure*}

Further downstream in the trade regions, Flowers make way to Gravel and Fish. These two patterns are more geographically intertwined, which is in agreement with the similarity of the environmental profiles in Fig.~\ref{fig:6}. Interestingly, Gravel seems to be relatively confined to the Barbados region, the west of Hawaii, and the southern tropical Pacific near regions -- like the South Pacific Convergence Zone --  of climatological convergence (Fig.~\ref{fig:2}). Hence the prevalence of Gravel in the more limited classification activity of \cite{Stevens2019a}, is not representative of the trade-wind regions more broadly. There is also some coincidence of Gravel hot-spots with regions of open-cell MCC regions as highlighted by the classification by \cite{Muhlbauer2014}, specifically around Hawaii and in the Southern sub-tropical Atlantic, but as with Flowers their MCC algorithm picks up many more open cells in higher latitudes. This, again, suggests that there may be a fundamental difference between the classes, something we already suspected based on their physical driving mechanisms, i.e. cold pools versus boundary layer circulations. Fish, appears linked to stronger synoptic upward motion (Fig.~\ref{fig:5}c), which the image snapshot from 1 May 2017 (Fig.~\ref{fig:6}a) suggests is associated with synoptic convergence lines, often connected to trailing mid-latitude fronts. 

Globally, the patterns are coherent, with hot-spots for a given pattern appearing in a few spatially extensive and plausibly similar meteorological regimes. This coherence supports the hypothesis that the subjective patterns are associated with meaningful and distinct physical processes. Though the combination of crowd-sourced labels and deep learning helped answering many of the questions raised at the outset of this study it also raises some new ones, for instance whether important cloud regimes are missing from our classification. Unsupervised classification algorithms like the one deployed by \cite{Denby2020} can be a good starting point to explore this question.

\section*{The Four Questions}

In this paper, we described a project to combine crowd-sourcing, to detect and label four subjectively defined patterns of mesoscale shallow cloud organization from satellite images, with deep learning.  The design and execution of the project raised a number of questions, four of which have been highlighted in this paper, and the answers to which we present as follows.


The first question (Q1) was concerned with how best to configure a crowd-sourcing activity. We found that speed and ease of use for the participants is paramount. Drawing crude rectangles on the screen only took tens of seconds for each image, whereas more detailed shapes such as polygons would have taken significantly longer. Further, the quickness of drawing boxes on an image meant that less of an attention span was required from the participants. (Some even reported to have had fun.) For our task, which involves judgements with inherent uncertainty, the added noise introduced by crude labels turned out to be insignificant in the statistical average, as shown by the ``consensus" found by the deep learning algorithm. Based on our experience, quantity trumps quality. This might, of course, be different for tasks where object boundaries are more clearly defined. 

Our second question, Q2, asked whether sufficient agreement exists between the human labelers to warrant scientific use of the labels. We believe that this is indeed the case. As discussed in the section titled "Inferences from human labels" there is a significant amount of disagreement between the participants, particularly because many cloud formations did not fit one of the four classes exactly. However, more importantly there was significant agreement on patterns that closely matched the canonical examples of ``Sugar'', ``Flower'', ``Fish'' and ``Gravel''. Taking a statistical average -- training a deep learning model can be viewed as doing just that -- removes some of the ambiguity from the labels and crystallizes the human consensus. Of course, the four classes chosen are not a complete description of all modes of organization, and others could have been defined. But the fact that the results are compatible with physical understanding suggest that the four classes do indeed capture important modes of cloud organization in the sub-tropics.

Q3 asked whether deep learning can be used to build an automated labeling system. The answer is a resounding yes. Both deep learning algorithms used in this paper, show high agreement scores. Further, visual analysis of the deep learning predictions suggest that these are less noisy than the human predictions. In other words, the deep learning models have learned to disregard the noise of the human labels and instead extract the common underlying pattern behind points of agreement, i.e., the essence of the proposed patterns. In addition, the deep learning models are both many orders of magnitude faster than humans at labeling images, and less costly and difficult to maintain.

The application of deep learning enabled us to classify a significantly larger geographical and temporal set of data.  This allowed us to look at  global patterns of ``Sugar'', ``Flower'', ``Fish'' and ``Gravel'' thereby addressing our fourth research question (Q4).  Here our main finding is that heat-maps of pattern occurrence are distributed in a geographically coherent way across all the major ocean basins, and sample significantly different meteorological conditions.  Heat maps for two of the patterns (Flower and Gravel) show some overlap with closed-cell mesoscale cellular convection, but only over portions of the sub-tropics.  As a rule the regions where patterns are identified (particularly for Fish and Flower) are not in regions familiar from past work on cloud classification.   

\section*{Inferences and Outlook}

The coherence of the heat maps for individual patterns suggests the presence of physical drivers underpinning their occurrence; drivers that may change as the climate changes.  Using the same classification categories but a different way of classifying the images, \cite{BonyEtAl2019} showed that differences in cloud radiative properties are associated with different forms of organization.  Our study thus lends weight to the idea that quantifying the radiative effects of shallow convection, and potential changes with warming, may require an understanding of, or at least ability to represent, the processes responsible for the mesoscale organization of fields of shallow clouds.   This might seem to be a daunting task. However, if the occurrence of different modes of organization can be reliably linked to large-scale conditions, reanalysis data or historical climate model simulations could help reconstruct cloud fields.  This could offer clues as to how meso-scale organization, and hence cloudiness, has changed in the past, and may change in the future.

This example helps highlight how crowd-sourced and deep-learned data-sets create new ways to study factors influencing shallow clouds and their radiative properties, and hopefully stimulates ideas for adapting the approach to other problems.  The growing accessibility of these new research methodologies makes their application all the more attractive.  Platforms like Zooniverse make it easy to set up a labeling interface free of charge. Plus, even if we did not do so, it is also possible to make the interface open to the public. Deep learning has also become much more accessible. Easy-to-use Python libraries\footnote{Keras \citep{Chollet2015} and fastai (\url{https://docs.fast.ai/}) were used for this study, see Supplement.} with pre-trained models for many applications in computer vision as well as accessible online courses\footnote{\url{https://course.fast.ai/}, \url{https://deeplearning.ai}} make it possible even for non-computer scientists to apply state-of-the art deep learning techniques. 

Our study also illustrates how crowd-sourcing and deep-learning effectively complement one another, also for problems in climate science. Deep learning algorithms typically need thousands of samples for training. These are not readily available for most problems in the geosciences.  A key lesson from our project is that, even for the ambiguously defined images that characterize many problems in atmospheric and climate science, it is feasible to create sufficient training data with a moderate amount of effort. We found that \num{5000} labels (i.e. a 6th of what was collected here) were enough to obtain similarly good results to the ones shown here. This translates to a day of labeling for around 15 people. 

This means that combining crowd-sourcing and deep learning is a promising approach for many questions in atmospheric science where features are easily -- albeit not unambiguously --  detectable by eye but hard to quantify using traditional algorithms. In our case, the combination of the two tools allowed us to generate global heatmaps, something that would have been impossible with traditional methods. Potential examples of similarly suited problems in the geosciences are detecting atmospheric rivers and tropical cyclones in satellite and model output\footnote{See \url{https://www.nersc.gov/research-and-development/data-analytics/big-data-center/climatenet/} for a similar project.}, classifying ice and snow particles images obtained from cloud probe imagery, or even large-scale weather regimes.

{\small
    \subsubsection*{Data availability}
    All data and code are available at \url{https://github.com/raspstephan/sugar-flower-fish-or-gravel}
    
    \subsubsection*{Acknowledgements}
    First and foremost, we would like to thank all the participants of the cloud labeling days. Special thanks go to Ann-Kristin Naumann and Julia Windmiller for initiating this collaboration and to Katherine Fodor for suggesting Zooniverse. SR acknowledges funding from the German Research Foundation Project SFB/TRR 165 ``Waves to Weather". This paper arises from the activity of an International Space Science Institute (ISSI) International Team researching ``The Role of Shallow Circulations in Organising Convection and Cloudiness in the Tropics''. Additional support was provided by the European Research Council (ERC) project EUREC4A (Grant Agreement 694768) of the European Union's Horizon 2020 Research and Innovation Programme and by the Max Planck Society. We acknowledge the use of imagery from NASA Worldview, part of the NASA Earth Observing System Data and Information System (EOSDIS).
    
    \subsubsection*{Preprint notice}
    This work has not yet been peer-reviewed and is provided by the contributing authors as a means to ensure timely dissemination of scholarly and technical work on a noncommercial basis. Copyright and all rights therein are maintained by the authors or by other copyright owners. It is understood that all persons copying this information will adhere to the terms and constraints invoked by each author's copyright. This work may not be reposted without explicit permission of the copyright owner.
}

{\footnotesize \bibliography{references}}
\clearpage
\end{multicols}

\section*{Supplemental Methods}
\renewcommand{\thefigure}{S\arabic{figure}}
\setcounter{figure}{0}
\renewcommand{\thetable}{S\arabic{table}}
\setcounter{table}{0}

\subsection{Region selection criteria}
The regions were  selected ahead of the classification days according to a similarity analysis of atmospheric conditions that resemble the conditions encountered during the DJF season east of Barbados where these patterns were first found \citep{Stevens2019a}.

Because the mesoscale organization of shallow cumulus is a relatively new research topic, the meteorological conditions influencing it are primarily an educated guess. Lower tropospheric stability (LTS), surface wind speed (FF) and total integrated column water vapour (TCWV) are three parameters one could naively imagine to describe the meteorological setting to a sufficient degree. Starting with the inter-annual seasonal mean of these atmospheric properties at the region east of Barbados, we searched for climatologically similar regions and seasons within a 120$^\circ$-wide latitudinal belt (60$^\circ$N to 60$^\circ$S) around the globe. We used a k-means clustering with eight clusters to find similar patterns within our search perimeter. As input to the algorithms we used the climatological means of LTS, FF10 and TCWV for each of the four seasons. The eight clusters explain more than 90\% of the variance in the dataset and provide large enough regions to fit 21$^\circ$ longitude by 14$^\circ$ latitude boxes reasonably well.

\begin{figure}[ht!]
	\centering
	\includegraphics[width=\textwidth]{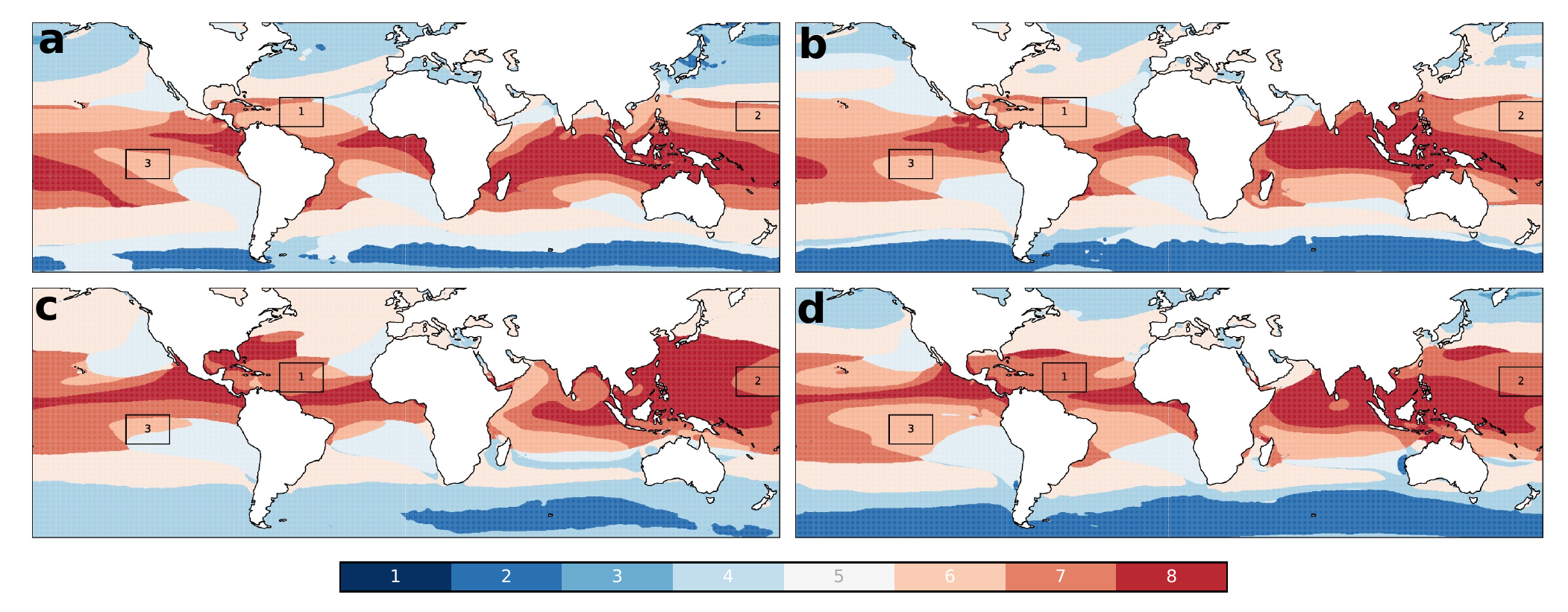}
	\caption{Cluster analysis of LTS, FF10, TCWV separated by season (DJF, MAM, JJA, SON). The colors identify the 8 clusters as a result of the k-means algorithm. For a better visual impression the clusters are sorted by cluster mean column integrated moisture with cluster 1 being the driest. Black boxes indicate regions chosen for human-classifications.}
	\label{fig:S1}
\end{figure}

Fig.~\ref{fig:S1} shows the clusters for the four seasons. Our analysis indicates that the meteorological conditions over the Northwestern Atlantic change with season. This is not surprising due to the migration of the ITCZ, but it illustrates that we shouldn't expect to see the same cloud patterns or at least the same distribution throughout the year. The final choice of seasons and regions was made to match the climate of region 1 in DJF (Table~\ref{tab:domain_description})

\begin{table}[ht!]
	\centering
	\caption{Selected domains used for human-classification of cloud patterns.}
	\label{tab:domain_description}
	\begin{tabular}{ccc}
		Domain&Bounds&Seasons used\\
		\hline
		1&-61$^\circ$E -40$^\circ$E; 10$^\circ$N 24$^\circ$N&DJF, MAM\\
		2&159$^\circ$E 180$^\circ$E; 8$^\circ$N 22$^\circ$N&DJF\\
		3&-135$^\circ$E -114$^\circ$E; -1$^\circ$N -15$^\circ$N&DJF, SON
	\end{tabular}
\end{table}

\subsection{Deep learning models}
Two deep learning models are used, one for object detection and one for semantic segmentation. For object detection, an algorithm called Retinanet \citep{Lin2017a} is used. Here we used the following implementation in Keras \citep{Chollet2015}: \url{https://github.com/fizyr/keras-retinanet}, which uses a Resnet50 \citep{He2015} backbone. The original images had a resolution of 2100 by 1400 pixels. For Retinanet the images were downscaled to 1050 by 700 pixels. This is necessary to fit the batch (batch size = 4) into GPU RAM.

For semantic segmentation, we first converted each human classification, i.e. all boxes by one user for an image, to a mask. Sometimes boxes for different patterns overlap. In this case, the mask is chosen to represent the value of the smaller box. Overall, the amount of overlapping boxes is small, however, so that the resulting error is most likely negligible. To create a segmentation model, we used the fastai Python library v1\footnote{\url{https://docs.fast.ai/}}. The network architecture has a U-Net \citep{Ronneberger2015a} structure with a Resnet50 backbone. For the segmentation model the images were downscaled to 700 by 466 pixels (batch size = 6).

To create the prediction masks, first a Gaussian filter with a half-width of 10 pixels was applied to smooth the predicted field. Then, for each pixel the highest probability for each of the four patterns was used, if this probability exceeded 30\%. This last step counteracts the tendency to predict background, which is by far the most common class in the training set.  

\subsection{Global heatmaps}

To create the heatmaps, the segmentation algorithms was used. Predictions were created for a 21$^\circ$ longitude by 14$^\circ$ latitude region at a time, with a windows sliding in 10.5$^\circ$ and 7$^\circ$ increments over the globe. The highest pattern probability for the overlapping images was then taken to create the global mask. This was necessary because the algorithm tends to predict background at the edges of the image, a consequence of the human labelers not drawing boxes that extend all the way to the edge of the image. The climatology was created from one year of Aqua data.
\newpage
\begin{figure}[th!]
    \centering
    \includegraphics[width=0.6\linewidth]{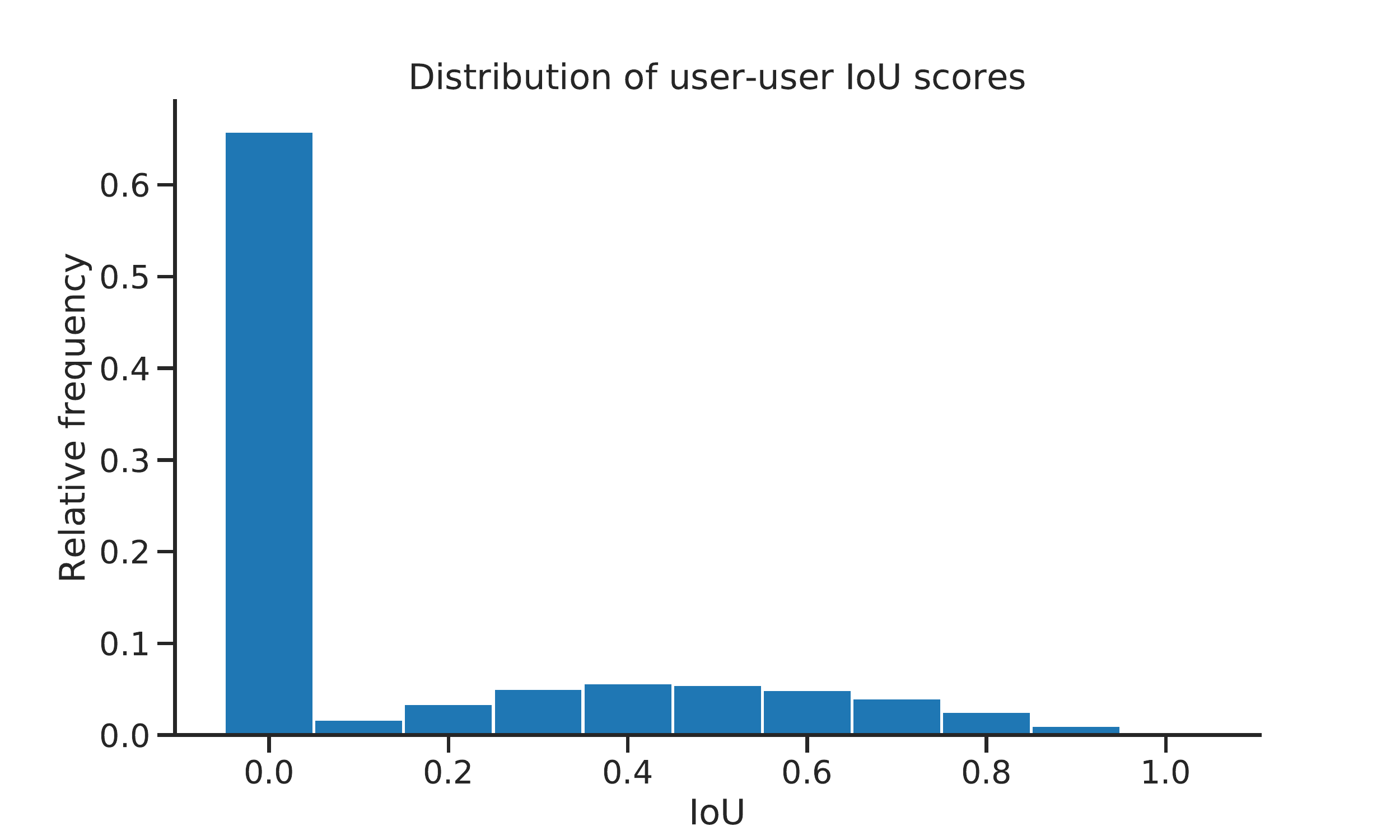}
    \caption{Histogram of IoU values for each user-user comparison for each class. Zero values mostly indicate cases where one user labeled a feature of a given class and the other user did not.}
    \label{fig:S2}
\end{figure}

\begin{figure}[th!]
    \centering
    \includegraphics[width=1\linewidth]{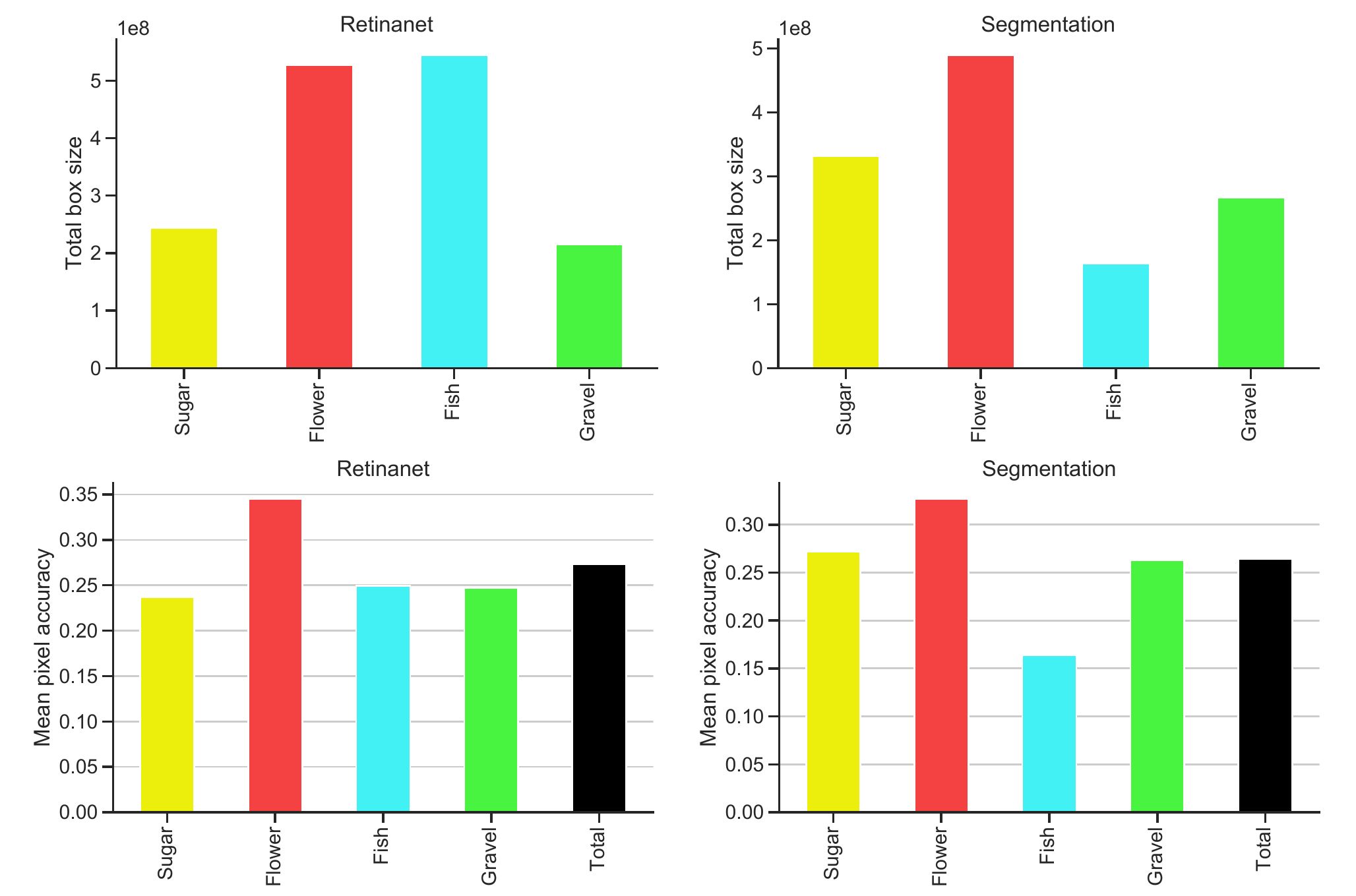}
    \caption{(Top row) Total size of classifications for the two deep learning algorithms for a random validation dataset. (Bottom row) Mean pixel accuracy (= mean IoU) for the two algorithms stratified by pattern, also for a random validation set.}
    \label{fig:S3}
\end{figure}
\newpage

\end{document}